\documentclass[%
 reprint,
superscriptaddress,
amsmath,amssymb,
aps,
]{revtex4-1}

\usepackage{graphicx}
\usepackage{amssymb}
\usepackage{dcolumn}
\usepackage{bm}
\usepackage{xcolor}
\usepackage{todonotes}
\usepackage{hyperref}
\hypersetup{
    colorlinks=true,
    linkcolor=blue,
    filecolor=magenta,      
    urlcolor=red,
    citecolor=blue,
}
\newcommand{\blue}{\textcolor{blue}}

\usepackage{xcolor}

\usepackage{soul}

\begin{document}
\title{Evolution of orbital excitations from insulating to superconducting MgTi$_2$O$_4$ films}

\author{Qizhi Li}\thanks{These authors contributed equally to this work.}
\affiliation{International Center for Quantum Materials, School of Physics, Peking University, Beijing 100871, China}

\author{Abhishek Nag}\thanks{These authors contributed equally to this work.}
\affiliation{Diamond Light Source, Harwell Campus, Didcot OX11 0DE, United Kingdom}

\author{Xiquan Zheng}\thanks{These authors contributed equally to this work.}
\affiliation{International Center for Quantum Materials, School of Physics, Peking University, Beijing 100871, China}

\author{Fucong Chen}
\affiliation{Beijing National Laboratory for Condensed Matter Physics, Institute of Physics, Chinese Academy of Sciences, Beijing 100190, China}

\author{Jie Yuan}
\affiliation{Beijing National Laboratory for Condensed Matter Physics, Institute of Physics, Chinese Academy of Sciences, Beijing 100190, China}

\author{Kui Jin}
\affiliation{Beijing National Laboratory for Condensed Matter Physics, Institute of Physics, Chinese Academy of Sciences, Beijing 100190, China}

\author{Yi Lu}
\email{yilu@nju.edu.cn}
\affiliation{National Laboratory of Solid State Microstructures and Department of Physics, Nanjing University, Nanjing 210093, China}
\affiliation{Collaborative Innovation Center of Advanced Microstructures, Nanjing University, Nanjing 210093, China}

\author{Ke-jin Zhou}
\email{kejin.zhou@diamond.ac.uk}
\affiliation{Diamond Light Source, Harwell Campus, Didcot OX11 0DE, United Kingdom}

\author{Yingying Peng}
\email{yingying.peng@pku.edu.cn}
\affiliation{International Center for Quantum Materials, School of Physics, Peking University, Beijing 100871, China}
\affiliation{Collaborative Innovation Center of Quantum Matter, Beijing 100871, China}

\date{\today} 

\begin{abstract}

Spinel oxides are well-known functional materials but rarely show superconductivity. Recently, emergent superconductivity was discovered in MgTi$_2$O$_4$, which is attributed to the increase of electron doping and the suppression of orbital order. Here, we utilized Ti $L$-edge resonant inelastic X-ray scattering to study the orbital excitations in superconducting (SC) and insulating MgTi$_2$O$_4$ films. 
We find that the spectral weight of orbital excitations is enhanced and the energy of $t_{2g}$ intra-band excitation is softened in the SC film compared to the insulating one, suggesting higher electron doping and suppressed orbital order gap in the SC sample. These observations were further supported by our multiplet calculations using minimal two-site model.
Our results provide spectroscopic evidence for the competition between orbital order and superconductivity in MgTi$_2$O$_4$ and shed light on searching for novel superconductors in spinel oxides.

\end{abstract}

\maketitle

Spinel compounds have attracted profound attention for their rich functionalities. Plenty of novel physical properties were revealed in these compounds, such as heavy fermion behaviors \cite{LVOHeavyFermion}, charge ordering \cite{AVOCO} and spin-Peierls transition \cite{ZCOSP}. However, for a long time, LiTi$_2$O$_4$ (LTO) was the only known superconductor among the spinel oxide family \cite{LTOSC} with a transition temperature (T$_c$) of $13$ K. The mechanism of superconductivity in these Ti-based superconductors are still heavily debated. Though being categorized as a BCS superconductor by specific heat \cite{LTOsh} and Andreev reflection spectroscopy \cite{LTOBCSswave}, an abnormally large upper critical field (H$_{c2}$) and an unconventional pseudogap were also observed by transport measurement \cite{LTOUpperField} and scanning tunneling spectroscopy \cite{LTOPG}. Previous resonant X-ray scattering experiments validated the relationship between SC and Ti$^{3+}$ d–d electron correlations in LTO \cite{Chen_2011}, which is analogous to the two-dimensional superconductivity in LaAlO$_3$/SrTiO$_3$ heterostructure \cite{LAO/STO}. 

MgTi$_2$O$_4$ (MTO) is a close relative of LiTi$_2$O$_4$, sharing the same crystal structure. Despite the similarity between the two compounds, MTO was not expected to become a superconductor because it displays a metal-to-insulator transition (MIT) at 260 K. This MIT is considered to be induced by an orbital order and spin-singlet dimerization, which is accompanied by a cubic-to-tetragonal lattice transformation \cite{MTOINS,MTOINS2,MTOINS3,MTOINS4,ZhuModern}. The Peierls state at low temperatures opens a symmetry-breaking gap at the Fermi surface, thus MTO is generally considered a robust Mott insulator \cite{Mottinsulator}. Recently, superconductivity has been realized in MTO films with elongated $c$-axis \cite{Weihuprb}, in analogy to the (Li,Fe)OHFeSe \cite{FeSe11111} and La$_{1.56}$Sr$_{0.44}$CuO$_4$/La$_2$CuO$_4$ superconductors \cite{LSCOhetero}. The emergence of superconductivity is attributed to the suppression of the orbital order and the increase of the density of states near Fermi energy according to density-functional calculations \cite{Weihuprb}. 

A microscopic mechanism of superconductivity involves various collective excitations, which are uncovered in different superconductors. For instance, the electron coupling to longitudinal optical (LO$_4$) phonon modes is proposed to play an essential role in the superconductivity of SrTiO$_3$ \cite{STOEPC}. The lattice excitations and electronic interaction are considered to enhance each other in a positive-feedback loop in cuprates \cite{yuhesci}. For orbital excitations, a systematic change of the \emph{dd} exciton spectrum has been observed in cuprate when the materials change from the normal state into the superconducting state \cite{ddPRX}.  In MTO, Raman and infrared reflectivity experiments revealed phonons at 40 $-$ 80 meV, plasma at 70 $-$ 120 meV and a temperature-dependent broad feature centered at $\sim$ 60 meV that is attributed to spin or polaronic excitations \cite{MTORaman}. 
The collective excitations relevant to SC in MTO are not clear yet. Whether the suppression of orbital order and the emergence of SC leave any traces in the orbital or the lower energy excitations remains an open question.

To identify the excitations in MTO and their connection to superconductivity, here we employ Ti $L$-edge resonant inelastic X-ray scattering (RIXS) to study superconducting (SC) and insulating (Non-SC) MTO thin films. As a result, we uncover phonons and orbital excitations including t$_{2g}$ intra-band and t$_{2g}-$e$_g$ inter-band orbital excitations in MTO, which is further supported by our multiplet calculations using minimal two-site model. We observed no discernable difference in phonons between SC and Non-SC samples. On the other hand, the intensities of the orbital excitations are stronger in the SC film, suggesting a relatively higher Ti$^{3+}$ concentration. More importantly, the energy of t$_{2g}$ intra-band excitation is largely softened in the SC film with respect to the Non-SC film, pointing to a suppressed orbital order gap.

\begin{figure}[htbp]
\centering\includegraphics[width = \columnwidth]{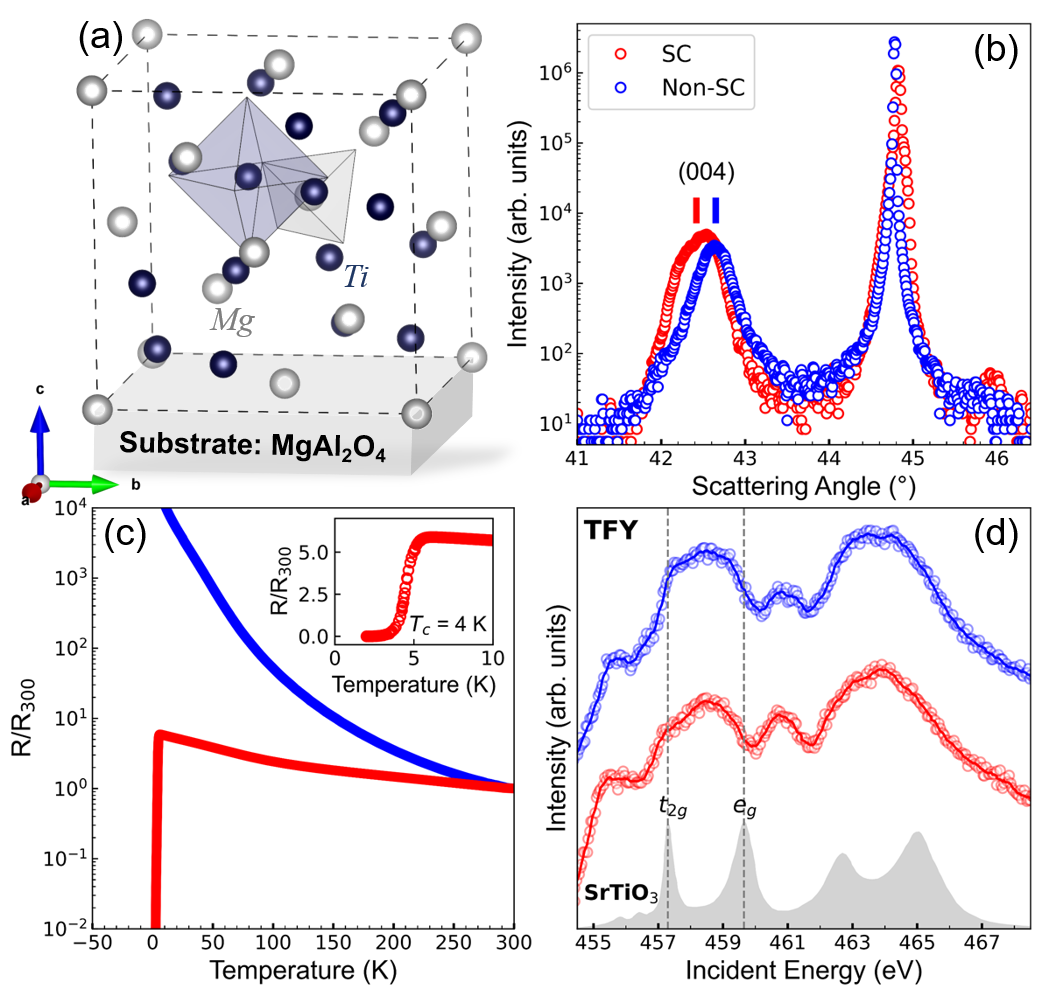}
\caption{\label{fig:characterization}
\textbf{(a)} The crystal structure of MgTi$_2$O$_4$ (MTO) on top of the substrate of MgAl$_2$O$_4$. We omit the oxygen atoms for clarity. \textbf{(b)} X-ray diffraction measurements for (004) Bragg peaks (intensity in log scale). The left peak arises from MTO films and the right peak from substrates. \textbf{(c)} The measurements of resistivity as a function of temperature. The transition temperature of the SC film is highlighted in the inset. \textbf{(d)} The TFY X-ray absorption spectra measured at Ti $L$-edge. We include the XAS of SrTiO$_3$ with Ti$^{4+}$ for comparison \cite{STOXAS}. A vertical offset is added for clarity.}
\end{figure}

The (00l)-oriented MgTi$_2$O$_4$ films were epitaxially grown on (00l)-oriented MgAl$_2$O$_4$ substrates (3 mm $\times$ 3 mm $\times$ 0.5 mm) by pulsed laser deposition (PLD). All samples were grown in a high vacuum of better than 1 $\times$ 10$^{-6}$ Torr with a pulse energy of $\sim$ 250 mJ and a repetition rate of 10 Hz. The deposition temperatures were 800$^\circ$C and 650$^\circ$C for SC and Non-SC MTO films, respectively. The thickness of MTO films was 500 nm (SC) and 200 nm (Non-SC) owing to the different deposition times. MTO is a cubic spinel oxide at room temperature with a three-dimensional network of corner-sharing tetrahedra formed by Ti$^{3+}$ cations as shown in Fig.~\ref{fig:characterization}(a). 
The oxygen atoms form octahedra and tetrahedra around Ti$^{3+}$ and Mg$^{2+}$, respectively. The dominant near-cubic crystal field splits the Ti$^{3+}$ d levels into t$_{2g}$ and e$_g$ ones \cite{MTOINS4}. We have characterized the films by X-ray diffraction (Fig.~\ref{fig:characterization}(b)) and resistivity measurements (Fig.~\ref{fig:characterization}(c)). The lattice constants are $a=b=8.52~\textrm{\AA}$ \cite{Weihuprb,MTOINS}, $c=8.468~\textrm{\AA}$ in the SC film and $c=8.458~\textrm{\AA}$ in the Non-SC film (\blue{see \emph{supplementary} Fig. S1}). 
The resistivity measurements show that the T$_c$ of SC film is $\sim$ 4 K, comparable with MgTi$_2$O$_4$/SrTiO$_3$ superlattice \cite{Weihuprb}.

\begin{figure} [htbp]
\centering\includegraphics[width = 0.96\columnwidth]{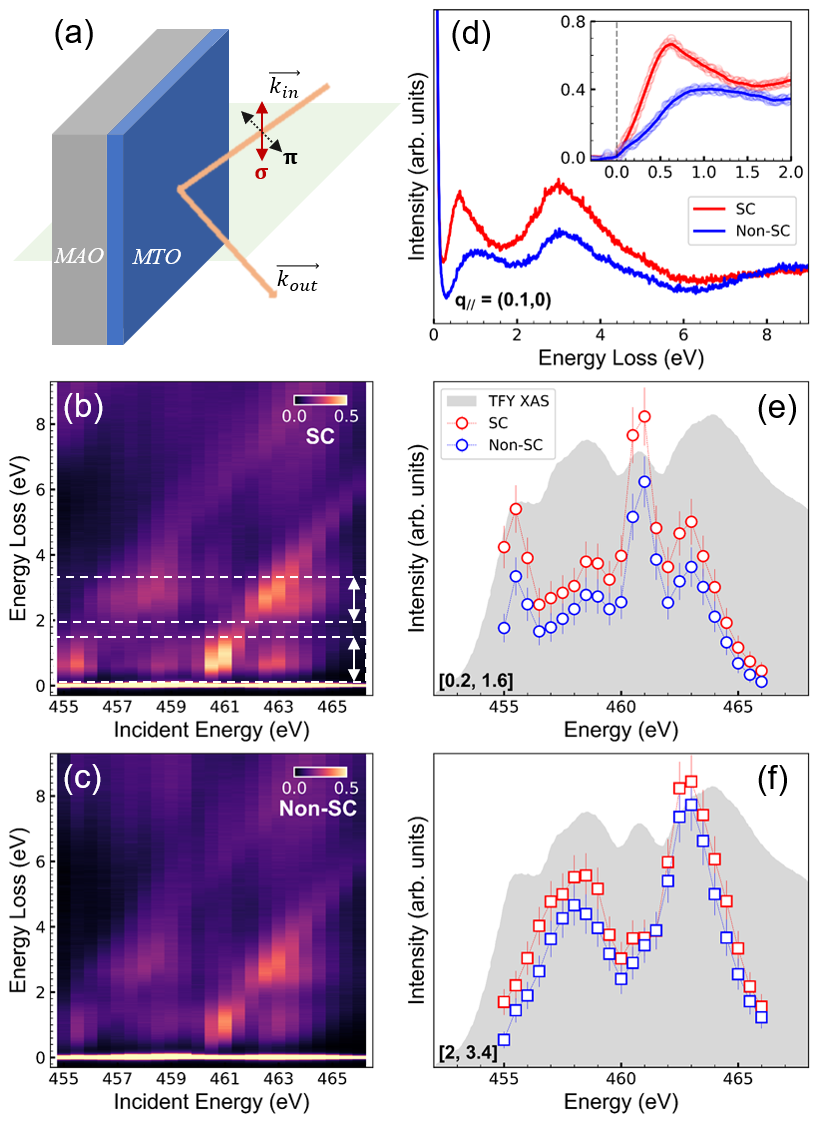}
\caption{\label{fig:detuning}
\textbf{(a)} RIXS experimental geometry. \textbf{(b,c)} RIXS intensity map versus energy loss and detuning energy across the Ti L-edge for SC and Non-SC MTO films at $\mathbf{q_\parallel}$ = $(0.1,0)$. The dotted lines and arrows indicate two Raman features. \textbf{(d)} The comparison of two representative RIXS spectra collected at 459 eV. The inset shows the low-energy spectra after subtracting the elastic peak and phonons. \textbf{(e,f)} The incident energy dependence of the integrated intensity within $[0.2,1.6]$ eV and $[2,3.4]$ eV energy windows indicated by arrows in panel (b) or (d). Error bars were estimated by the noise level of the spectra. 
The XAS of SC film is overlaid for comparison.}
\end{figure}

We measured the Ti $L$-edge X-ray absorption spectrum (XAS) at normal incidence with the total fluorescence yield (TFY) method (Fig.~\ref{fig:characterization}(d)). The nominal valance of Ti in MTO is Ti$^{3+}$, but there is some Ti$^{4+}$ due to surface oxidation by comparing with the XAS of SrTiO$_3$ \cite{STOXAS}. The dominating distribution of Ti$^{3+}$ is manifested by the TFY channel and consistent with the previous electron energy loss spectroscopy (EELS) results \cite{Weihuprb}.  
The XAS and RIXS spectra were collected at the I21 beamline of the Diamond Light Source \cite{I21beamline}. The energy resolution for RIXS spectra ranged from 22 to 40 meV by measuring the width of the elastic line on a carbon tape. All the experiments were performed at base temperature $\sim$ $12$ K. The scattering plane was horizontal, defined by the incident and outgoing beams with the geometry shown in Fig.~\ref{fig:detuning}(a). The scattering angle was fixed at 150$^\circ$, so the maximum accessible in-plane momentum transfer ($\mathbf{q_\parallel}$) was 0.61 $r.l.u.$ in the reciprocal space. Unless otherwise specified, the RIXS spectra were collected with $\sigma$-incident polarization.

We first performed the incident-energy detuning measurements to distinguish the excitations between Raman features and the fluorescence background. The Raman features are localized excitations, whose energies are independent of the detuning energy, while the fluorescence background comes from itinerant carriers or continuum, increased with incident photon energy \cite{LAO/STO,Chen_2011,2.5dd}. Figure ~\ref{fig:detuning}(b,c) display a series of RIXS spectra across Ti L-edge in steps of 0.5 eV for SC and Non-SC films, respectively. 
Below the energy loss of 4 eV, the RIXS intensity maps show two broad Raman features. 
To understand their origins we select two energy windows of $[0.2,1.6]$ eV and $[2,3.4]$ eV indicated by two arrows in Fig.~\ref{fig:detuning}(b)) and track their intensities with incident energy. To facilitate the comparison of the two samples, we normalized the RIXS spectra to their high-energy background  ($\sim[6,9]$ eV), which overlapped for the two films as shown in Fig.~\ref{fig:detuning}(d). The integrated intensities within the two energy windows are shown in Fig.~\ref{fig:detuning}(e,f) for both samples. The $[0.2,1.6]$ eV region resonates at Ti$^{3+}$ t$_{2g}$ absorption peak ($\sim$ 455.5 eV and 461 eV), suggesting a t$_{2g}$ intra-band transition. The $[2,3.4]$ eV region resonates at Ti$^{3+}$ e$_{g}$ absorption peak ($\sim$ 458.5 eV and 463 eV), suggesting a t$_{2g}$ to e$_{g}$ inter-band transition (10Dq) \cite{STOXAS,2.5dd}. Figure~\ref{fig:detuning}(d) displays stronger \emph{dd} orbital excitations in SC samples than Non-SC ones. Since the 3$d$ orbital is empty in Ti$^{4+}$ (3$d^0$) and does not contribute to \emph{dd} orbital excitations, the orbital excitation intensity can only arise from Ti$^{3+}$ (3$d^1$) concentration \cite{STOXAS,2.5dd}. Therefore, the stronger \emph{dd} orbital excitations in SC film indicate more Ti$^{3+}$ cations. The inset of Fig.~\ref{fig:detuning}(d) shows the low-energy region after the subtraction of elastic peak and phonons (\blue{see \emph{supplementary} Fig. S2}). It is clear that the spectral weight shifts to higher energy in Non-SC film indicating the existence of an energy gap, which was reported as $\sim$ 0.25 eV due to orbital ordering in bulk MTO from optical measurements \cite{MTOgap}. 

\begin{figure}[htbp]
\centering\includegraphics[width = \columnwidth]{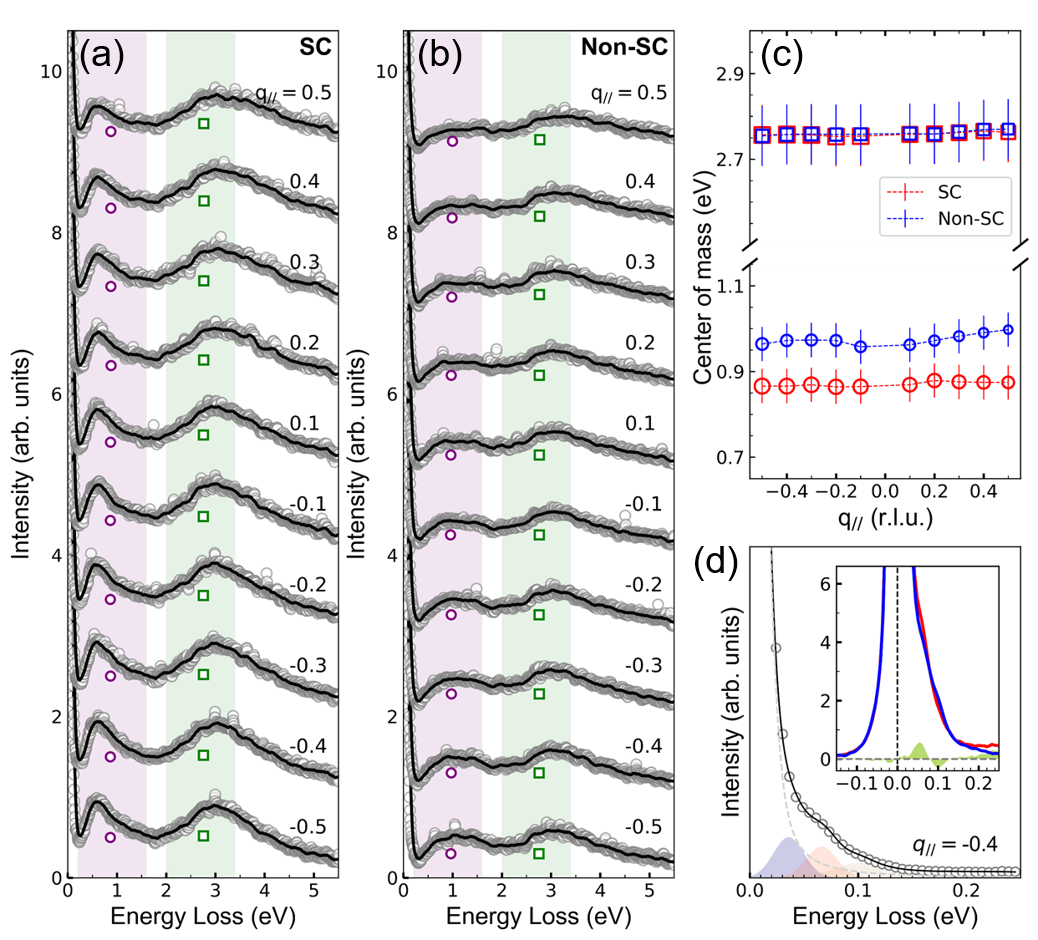}
\caption{\label{fig:qdep}
\textbf{(a,b)} RIXS spectra for SC and Non-SC MTO films collected at 458.5 eV with $\mathbf{q_\parallel}$ indicated next to the curves. Each spectrum is shifted vertically for clarity. The shaded areas indicate the energy windows, and the COM are marked by circles and squares. \textbf{(c)} Momentum dependence of the COM in two energy windows for SC and Non-SC MTO films. The sizes of the markers indicate the integrated intensity. Error bars were estimated via the uncertainty of the energy windows. \textbf{(d)} The fitting of phonons. The inset compares the low energy region for SC and Non-SC MTO samples.}
\end{figure}

Then we performed momentum-dependent measurements along $\mathbf{q_\parallel}$ = (H,0) direction at the incident energy of 458.5 eV. 
The RIXS spectra are shown in Fig.~\ref{fig:qdep}(a,b), and the shaded areas represent two groups of \emph{dd} excitations, each of which consists of multiple peaks. We calculated the center of mass (COM) as their energies in Fig.~\ref{fig:qdep}(c), and the size of the markers indicates the intensities. The dispersionless behavior confirms the local nature of \emph{dd} orbital excitations from Ti$^{3+}$. The intensities of \emph{dd} excitations in SC film are stronger than those of Non-SC film. Moreover, we observe that the energy of t$_{2g}$ intra-band transition softens by $\sim$ 0.1 eV in SC film compared with the Non-SC one, while the energies of 10Dq do not change between the two samples. We do not observe any dispersive excitations, suggesting the absence of magnetic or plasmon excitations in MTO \cite{Paramagnon,NiMagnon,LCCOplasma}.  

We can also explore the low-energy phonons in SC and Non-SC MTO films. Previous Raman spectroscopy suggests the existence of two stronger phonon modes (F$_{2g}$ $\sim$ 42 meV and F$_{2g}$ $\sim$ 62 meV) and two weaker phonon modes (E$_{g}$ $\sim$ 56 meV and A$_{1g}$ $\sim$ 79 meV) \cite{MTORaman}. Due to the limited energy resolution, we only fit the low-energy spectra with two resolution-limited Gauss peaks corresponding to the stronger F$_{2g}$ phonons and one anti-Lorentz peak for the higher energy phonon (A$_{1g}$ $\sim$ 79 meV) (\blue{see \emph{supplementary} Fig. S3}). One example of the fitting curve is shown in Fig.~\ref{fig:qdep}(d), and the inset shows the comparison of the two samples. We find there is no distinction for both phonon energy and intensity which relates to electron-phonon coupling (EPC) \cite{Ament_2011} between SC and Non-SC films. We suspect that the absence of any enhancement of EPC in SC film may relate to the phonon modes responsible for the SC not being captured or activated in RIXS measurements or the electronic correlation being more dominant in the emergence of SC in MTO.

\begin{figure} [htbp]
\centering\includegraphics[width = \columnwidth]{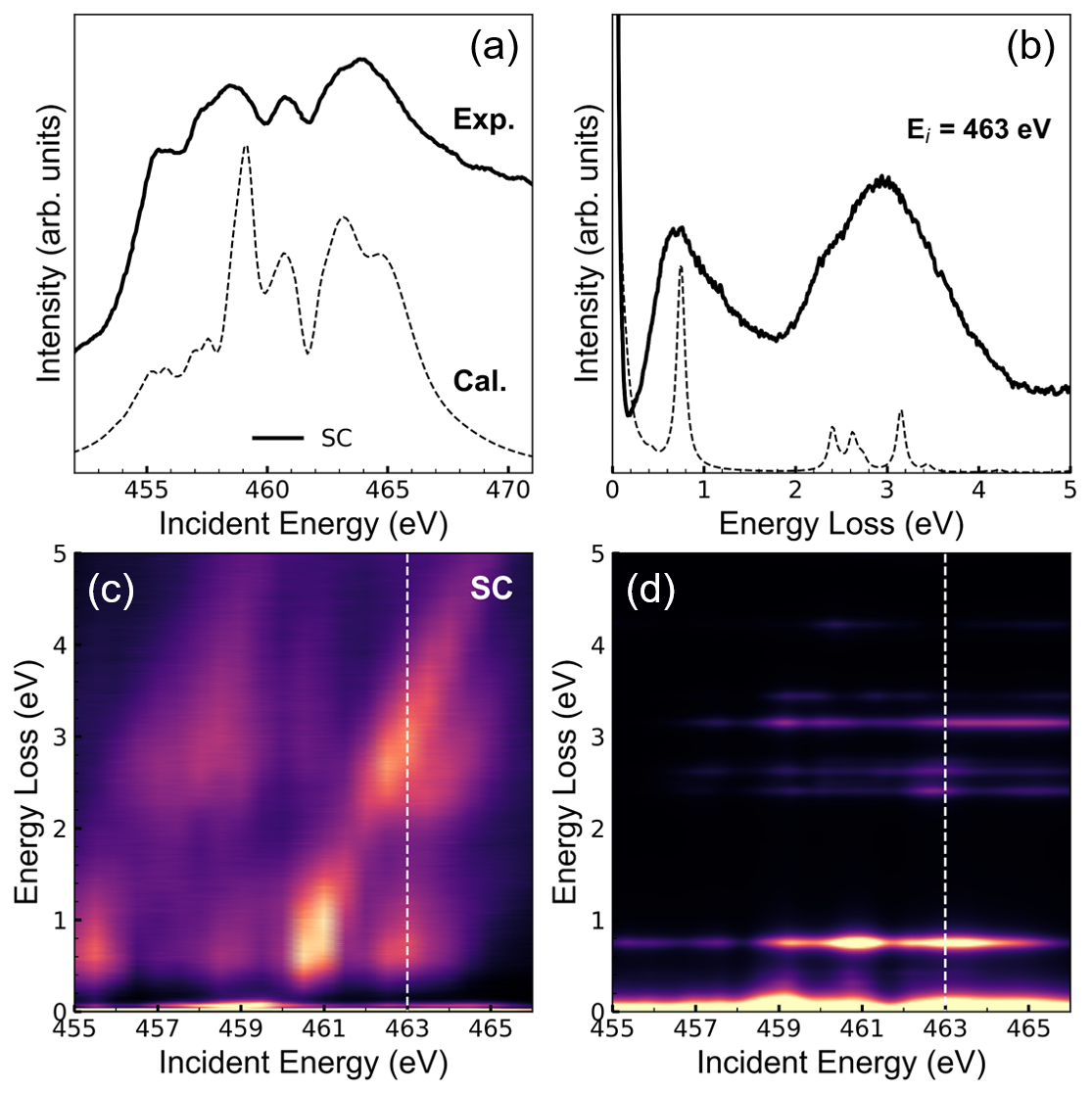}
\caption{\label{fig:model}
\textbf{(a)} Comparison of TFY XAS and \textbf{(b)} RIXS spectrum between experimental data and calculations with the two-site model on SC film, respectively. Comparison of detuning RIXS map between experiment \textbf{(c)} and calculation \textbf{(d)} on SC film. The selected RIXS spectrum in panel (b) is indicated by the dotted white line.}
\end{figure}

To further validate the interpretation of our experimental results and to gain more insights into the electronic structure of MTO, multiplet calculations were performed using \textsc{Quanty} \cite{Quanty,Haverkort_2016}. In order to capture the nonlocal band effects of the d orbitals, we adopted a minimal two-site model including the full d orbitals of two Ti$^{3+}$ ions with intersite hybridizations. Note that the local point group of the Ti$^{3+}$ ions in MTO is $D_{3d}$, which lifts the $t_{2g}$ degeneracy to $a_{1g}$ and $e_{g}'$. Density-functional calculations \cite{PhysRevB.78.125105} have shown that the splitting of on-site energies for these two energy levels is negligible without orbital order, which is thus left out in our calculation to keep the number of parameters minimal. The inter-site hoppings are set to $V_{a1g}=0.90$, $V_{eg}=0.72$, $V_{eg'}=0.72$ for the SC case. The $e_g$-$t_{2g}$ splitting $10Dq$ is set to 2.4. The 3d-3d and 2p-3d Coulomb interactions are set to $U_{dd}=3.5$ and $U_{pd}=4.0$, respectively, and the Slater integrals are scaled to 80\% of their atomic Hartree-Fock values \cite{Haverkortthesis}. All energy parameters are in units of eV.

Figure~\ref{fig:model}(a) shows the calculated XAS spectra with an energy-dependent Lorentzian broadening to account for decay channels not included in the model. The calculated spectral lineshape show overall good agreement with their experimental counterparts. 
Most notably, the calculation reproduces the small peak at around 461 eV that we have previously identified as absorption into the Ti$^{3+}$ $t_{2g}$ levels. Figure~\ref{fig:model}(b) shows the calculated RIXS single spectra in comparison to the experimental one. 
The energies of orbital excitations are in good agreement with the experiment. The stronger spectral weight between 2 $\sim$ 4 eV in the experiment is due to the contamination of the fluorescence background, which is omitted in the calculation. The calculated detuning RIXS map in Fig.~\ref{fig:model}(d) also agrees well with the experimental one in Fig.~\ref{fig:model}(c), especially the predominant resonance of the low-energy \emph{dd} excitation around 461 eV. The consistency between the calculation and the experiments for both XAS and RIXS lends support to our interpretation of the experimental results.

Previous energy-dispersive x-ray (EDX) measurements on MTO films showed that the ratio of Mg and Ti (Mg/Ti) decreases with higher deposition temperature \cite{Weihuprb}. With a reduced Mg/Ti ratio, the MTO films have higher electron doping and a larger $c$ lattice constant. Consequently, the orbital order is suppressed, as inferred from the absence of kink in the temperature-dependent transport measurement, and superconductivity emerges from this strongly correlated background. 
Here we observed a consistent increase of out-of-plane lattice from Non-SC to SC films (from 8.458 to 8.468~\textrm{\AA}). Furthermore, we observe strongly suppressed RIXS low-energy ($\lesssim$ 0.5 eV) spectral weight in the Non-SC film, indicating the existence of a gap. This also shifts the COM of the intra-$t_{2g}$ excitation up by 0.1 eV. We emphasize that such a large shift is unlikely the result of increased crystal-field splitting within the $t_{2g}$ orbitals since the difference of $c$ lattice constants between two films is only 0.12$\%$. A simple estimation using Harrison's rule \cite{PhysRevB.28.4363} would result in a change of less than 1\%.
And note that the resistance measurements of the Non-SC film show no clear kink feature (Fig.~\ref{fig:characterization}(c)), possibly due to a partially suppressed orbital order and/or structural relaxation.

In conclusion, we have measured the orbital excitations and phonons in superconducting and insulating MgTi$_2$O$_4$ films as a function of incident energy and transfer momentum. The phonons are nearly identical in the two films, while the orbital excitations are enhanced and intra-t$_{2g}$ band excitation is softened in the SC film. The difference highlights the important roles played by electron doping and suppression of orbital order in the appearance of superconductivity. The two-site calculations reproduce well the orbital excitations in the RIXS spectra of SC film. We notice that V-doped bulk Mg$_{1-x}$Ti$_2$O$_4$ has been recently reported as a superconductor with an increase of T$_c$ up to 16 K \cite{MVTO}. It is also intriguing to figure out whether the charge and orbital fluctuations introduced by V-doping assist in boosting the superconductivity.

\vspace{1 ex}
\begin{acknowledgments}
\noindent
The RIXS experimental data were collected at beamline I21 of the Diamond Light Source in Harwell Campus, United Kingdom. \textbf{Funding:} Y.~Y.~P. is grateful for financial support from the Ministry of Science and Technology of China (2019YFA0308401 and 2021YFA1401903) and the National Natural Science Foundation of China (11974029). Y.L. acknowledges support from the National Key R$\&$D Program of China under grant No. 2022YFA1403000 and the National Natural Science Foundation of China under grant No.12274207.
\end{acknowledgments}

\end{document}